# CodeTrans: Towards Cracking the Language of Silicon's Code Through Self-Supervised Deep Learning and High Performance Computing

**Ahmed Elnaggar**[*]                                                                                  AHMED.ELNAGGAR@TUM.DE
**Wei Ding**[*]                                                                                                   WEI.DING@TUM.DE
*Department of Informatics*
*TUM (Technical University of Munich)*
*Boltzmannstrasse 3, 85748 Garching, Germany*

**Llion Jones**                                                                                                    LLIONS@GOOGLE.COM
*Google AI*
*Google*
*1600 Amphitheatre Parkway, Mountain View, CA 94043, USA*

**Tom Gibbs**                                                                                                     TGIBBS@NVIDIA.COM
**Tamas Feher**                                                                                                   TFEHER@NVIDIA.COM
**Christoph Angerer**                                                                                         CANGERER@NVIDIA.COM
*Nvidia*
*2788 San Tomas Expy, Santa Clara, CA 95051, Vereinigte Staaten, USA.*

**Silvia Severini**                                                                                      SILVIA@CIS.UNI-MUENCHEN.DE
*Center for Information and Language Processing*
*Ludwig-Maximilians-Universität München*
*Oettingenstraße 67, D-80538 München, Germany.*

**Florian Matthes**                                                                                                MATTHES@TUM.DE
**Burkhard Rost**                                                                                       ASSISTANT@ROSTLAB.ORG
*Department of Informatics*
*TUM (Technical University of Munich)*
*Boltzmannstrasse 3, 85748 Garching, Germany*

## Abstract

Currently, a growing number of mature natural language processing applications make people's life more convenient. Such applications are built by source code — the language in software engineering. However, the applications for understanding source code language to ease the software engineering process are under-researched. Simultaneously, the transformer model, especially its combination with transfer learning, has been proven to be a powerful technique for natural language processing tasks. These breakthroughs point out a promising direction for process source code and crack software engineering tasks. This paper describes CodeTrans — an encoder-decoder transformer model for tasks in the software engineering domain, that explores the effectiveness of encoder-decoder transformer models for six software engineering tasks, including thirteen sub-tasks. Moreover, we have investigated the effect of different training strategies, including single-task learning, transfer learning, multi-task learning, and multi-task learning with fine-tuning. CodeTrans outperforms the state-of-the-art models on all the tasks. To expedite future works in the software engineering domain, we have published our pre-trained models of CodeTrans.[1]

---

∗. Equal contribution. Correspondence to ahmed.elnaggar@tum.de.

## 1. Introduction

Software engineering can be considered a process of designing, implementing, testing, and maintaining information systems such as applications, frameworks, or other software components (Royce, 1987) — this is a highly complex undertaking. So experienced specialists invent and use different tools and methods (for example, design patterns, code documentation, unit tests, version control tools, etc.) to control and improve the software quality and make the software engineering process more effective and convenient. In software engineering, works are done by using different programming languages. Programming language can be considered as a type of language to communicate with the computer systems to build the applications and achieve the requirements. Therefore, natural language processing techniques could also be applied to solve the programming languages' tasks to assist the software engineering process.

Recently, using models like BERT (Devlin et al., 2018), XL-NET (Yang et al., 2019), ALBERT (Lan et al., 2019), RoBERTa (Liu et al., 2019b), GPT-3 (Brown et al., 2020), and T5 (Raffel et al., 2020) have become a trend in natural language processing. All these models have the Transformer architecture of Vaswani et al. (2017) with attention mechanisms. Furthermore, their large pre-trained checkpoints are very suitable to be fine-tuned on downstream supervised tasks. Such a transfer learning technique helps transfer the knowledge gained from different large datasets to a small specific dataset, avoids overfitting, and saves computational cost. Among them, Raffel et al. (2020) carried out a large number of experiments to explore transfer learning performance and proved its effectiveness in natural language processing tasks. In addition, the Multi-Task Deep Neural Network (MT-DNN) proposed by Liu et al. (2019a) based on Bert also obtained excellent results on Natural Language Understanding tasks. MT-DNN involves the pre-training stage of Bert together with a multi-task learning approach and shows the latter's excellent performance.

In the software engineering domain, transformer models with transfer learning and multi-task learning are under-explored. According to the recent research by Watson (2020), the Recurrent Neural Network (Kombrink et al., 2011) (including long short-term memory (LSTM) (Hochreiter and Schmidhuber, 1997) and Gated Recurrent Unit (GRU) (Cho et al., 2014)) were the most popular model for software engineering tasks from 2000 to 2019. These kinds of models can understand a sequence of texts. Iyer et al. (2016) proposed CODE-NN with the architecture consisting of LSTM guided by a global attention model (Luong et al., 2015) to summarize SQL and CSharp code snippets. Hu et al. (2018) published the DeepCom composed by encoder-decoder LSTMs to generate code comments for Java code functions automatically. Jiang et al. (2017) developed the Neural Machine Translation (NMT) with encoder-decoder RNNs to generate commit messages from git change diffs. The model DeepAPI from Gu et al. (2016) to generate API usage sequences for a given natural language query is composed of attention-based encoder-decoder GRUs. Polosukhin and Skidanov (2018) also used GRU and RNN cells to build a model synthesizing LISP-inspired domain-specific language (DSL) code.

---

[1]. https://github.com/agemagician/CodeTrans

Nevertheless, the transformer models have also lately gained attention for software engineering tasks. Feng et al. (2020) proposed a transformer model called CodeBERT to support natural language and programming language tasks like natural language code search, code documentation generation, etc. Moreover, Lachaux et al. (2020) published a Transcoder model for translating functions between C++, Java, and Python based on the transformer architecture. These models all achieved state-of-the-art results for software engineering tasks.

This paper proposes our CodeTrans models for software engineering tasks based on the encoder-decoder transformer architecture. We applied CodeTrans to six different kinds of tasks, including Code Documentation Generation, Source Code Summarization, Code Comment Generation, Git Commit Message Generation, API Sequence Recommendation, and Program Synthesis. Furthermore, we trained CodeTrans using single-task learning, transfer learning, and multi-task learning on one NVIDIA GPU and Google Cloud TPUs. Also, we used both supervised tasks and self-supervised tasks for building a language model in the software engineering domain. Our models achieve state-of-the-art performance on all the tasks. We have also contributed our pre-trained checkpoints and published the models for each task in Hugging Face model Hub.[2]

The structure of this paper is as follows: In Section 2, we introduce the tasks of our experiments, the datasets, and our preprocessing methods. We explain the model architecture and the vocabulary representation in Section 3. We list our experiment details during the training in Section 4. The models' performances are compared in Section 5. In Section 6 we offer our reflection and conclusion. Future work necessary for further research is also introduced in this section.

## 2. Tasks and Datasets

This section describes the supervised tasks and the corresponding datasets, and the unlabeled datasets for self-supervised tasks. Furthermore, we explain the data preprocessing methods for different languages.

### 2.1 Supervised Tasks and Datasets

In this work, we trained six supervised tasks in the software engineering domain as follows.

**Code Documentation Generation:** For the Code Documentation Generation supervised tasks, a code function was an input to the model, and the model generated the documentation for this function. CodeSearchNet Corpus Collection[3] (Husain et al., 2019) was selected as the dataset for this task. It contains six programming languages' functions/methods, including Python, Java, Go, Php, Ruby, and Javascript. These functions with their documentation were downloaded from the Github[4] repository. We used the pre-processed version by CodeBERT[5] (Feng et al., 2020), which already provides a parsed and tokenized dataset.

---

2. `https://huggingface.co/models?search=code_trans`
3. `https://github.com/github/CodeSearchNet`
4. `https://github.com/`

**Source Code Summarization:** Given a short code snippet, the Source Code Summarization task generates a summary for this code. This task involves Python, SQL, and CSharp languages. We used the dataset[6] generated by Iyer et al. (2016). This dataset is extracted from StackOverflow.[7] The code snippets are from accepted answers that contain exactly one code snippet, and the summarization is the corresponding title of the question. Iyer et al. (2016) asked human annotators to provide two additional titles for 200 randomly chosen code snippets from the validation and test set for SQL and CSharp code. We followed their preprocessing methods and evaluation using the test dataset annotated by human annotators.

**Code Comment Generation:** Like Code Documentation Generation, this task focuses on generating the JavaDoc for the Java functions. We used the corpus[8] from Hu et al. (2018) for this task. They focused on Javadoc comments from 9,714 Java open source projects from Github. The first sentence in the Javadoc description is extracted as the expected comment.

**Git Commit Message Generation:** The task Git Commit Message Generation aims to generate a commit message describing the git commit changes. The development-assisted tool Git is a version-control system for tracking changes in files and codes during software engineering. A well-structured code commit helps to overview the project development and control the code changes and the development quality. We used the dataset[9] from Jiang et al. (2017) based on 1,000 Java repositories having the most Github stars.

**API Sequence Recommendation:** We aimed to generate the API usage sequence, including the class and function names, by giving the model a natural language description as an input. It would be beneficial to suggest to developers the API sequence when searching and asking about the corresponding usages. The dataset[10] was extracted by Gu et al. (2016) from Java projects with at least one star from Github.

**Program Synthesis:** Program synthesis is the task of synthesizing or generating programming codes based on natural language description. We used the AlgoLisp dataset[11] (Polosukhin and Skidanov, 2018) for the Program Synthesis task. This dataset was extracted from homework assignments for introductory computer science courses, so each example in this dataset consisted of a question and an answer. We input the question into our model and expect the model to output the correct LISP-inspired DSL code answer.

Table 1 compares the number of samples in training, validation, and testing datasets for each supervised learning task. We observe that the API Sequence Recommendation has the most significant number of samples. The second-largest dataset is the Code Comment Generation dataset. Four out of six tasks have the datasets extracted from GitHub. Furthermore, three out of six tasks used the function-level as an input rather than a complete program-level input.

---

5. `https://github.com/microsoft/CodeBERT`
6. `https://github.com/sriniiyer/codenn`
7. `https://stackoverflow.com/`
8. `https://github.com/xing-hu/DeepCom`
9. `https://sjiang1.github.io/commitgen/`
10. `https://github.com/guxd/deepAPI`
11. `https://github.com/nearai/program_synthesis/tree/master/program_synthesis/algolisp`

| Task | Language | Train | Valid | Test | Data Source | Data Level |
|---|---|---|---|---|---|---|
| Code Documentation Generation | Python | 251,820 | 13,914 | 14,918 | GitHub | Function |
| | Java | 164,923 | 5,183 | 10,955 | | |
| | Go | 167,288 | 7,325 | 8,122 | | |
| | Php | 241,241 | 12,982 | 14,014 | | |
| | Ruby | 24,927 | 1,400 | 1,261 | | |
| | Javascript | 58,023 | 3,885 | 3,291 | | |
| Source Code Summarization | Python | 12,004 | 2,792 | 2,783 | StackOverflow | Code Snippet |
| | Csharp | 52,943 | 6,601 | 108 | | |
| | SQL | 25,671 | 3,326 | 100 | | |
| Code Comment Generation | Java | 470,451 | 58,811 | 58,638 | GitHub | Function |
| Git Commit Message Generation | Java | 26,208 | 3,000 | 3,000 | GitHub | Commit |
| API Sequence Recommendation | Java | 7,475,850 | - | 10,000 | GitHub | API |
| Program Synthesis | DSL | 79,214 | 10,819 | 9,967 | Homework | Function |

Table 1: The summarization of supervised datasets. It includes the number of samples in training, validation and testing data-sets, data source, and the programming data level for each supervised dataset. Each sample could be a code function, a code snippet, a commit diff, or a natural language sentence, depending on the data level.

## 2.2 Unlabeled Datasets

Unlabeled datasets are used in the transfer learning pre-training stage and the multi-task learning. They are helpful when building a language model for tasks in the software engineering domain and make the final model more generalized against overfitting (Ruder, 2017). We involved six different corpora for unlabeled datasets, covering nine programming languages and English.

**CodeSearchNet Corpus Collection**: The CodeSearchNet Corpus Collection (Husain et al., 2019) was used as one of the supervised tasks and as one of the self-supervised tasks. This dataset is divided into two parts — functions *with* the function documentation and functions *without* documentation. In the self-supervised training, we have used both of these parts together. For functions without documentation, we directly considered each function as a separate sentence sequence. For functions with documentation, we concatenated each pair of tokenized-function and its tokenized-documentation as one input sentence sequence. All the code samples in this corpus are function-level samples.

**Public Git Archive:** We used CSharp and an additional Java datasets from the Public Git Archive dataset[12] (Markovtsev and Long, 2018). The Public Git Archive has code in the file-level containing the import statements, multiple functions, and comments. Such file-level data could help the models to understand more information like API usage and the relationship between different functions.

**150k Python Dataset:** We included the 150k Python Dataset[13] (Raychev et al., 2016) from the SRILAB at ETH Zurich. The Python programs in this data-set are collected from GitHub repositories with permissive and non-viral licenses by removing duplicate files, forked projects, and obfuscated files. The Python code in this corpus is also a file-level code like Public Git Archive datasets.

---
12. https://github.com/src-d/datasets/tree/master/PublicGitArchive
13. https://www.sri.inf.ethz.ch/py150

| Language | CodeSearchNet Without Doc | CodeSearchNet With Doc | 150k Python Dataset | The Public Git Archive | StaQC | LISP | One Billion Word Corpus | Total |
|---|---|---|---|---|---|---|---|---|
| Python | 657,030 | 375,210 | 149,114 | | | | | 1,181,354 |
| Java | 1,070,271 | 373,412 | | 720,124 | | | | 2,163,807 |
| Go | 379,103 | 300,882 | | | | | | 679,985 |
| Php | 398,058 | 369,923 | | | | | | 767,981 |
| Ruby | 110,551 | 43,803 | | | | | | 154,354 |
| Javascript | 1,717,933 | 99,646 | | | | | | 1,817,579 |
| CSharp | | | | 469,038 | | | | 469,038 |
| SQL | | | | | 133,191 | | | 133,191 |
| LISP | | | | | | 122,602 | | 122,602 |
| English | | | | | | | 30,913,716 | 30,913,716 |
| Total | | 5,895,822 | 149,114 | 1,189,162 | 133,191 | 122,602 | 30,913,716 | 38,403,607 |

Table 2: The number of samples of each unlabeled data-set for different programming languages and the English natural language. The first column lists the languages. For programming languages, each sample can be considered as one function or a programming file, or part of the code, depending on the code level of that dataset. For the English language, one sample means one sentence.

**StaQC:** For the SQL unlabeled dataset, we chose StaQC[14] (Yao et al., 2018). This dataset was extracted from StackOverflow. StaQC contained SQL question-code pairs of questions tagged by "SQL," "database," or "oracle" from StackOverflow. The SQL code in StaQC is code-snippet level and not the entire file-level code.

**LISP:** We created a new LISP dataset, with which we extracted 20 GitHub repositories[15] having the most stars from the Lisp Topic by applying the GitHub Rest API[16] and parsed the files into function-level LISP code.

**One Billion Word Language Model Benchmark Corpus:** Despite the programming languages, we used one Billion Word Language Model Benchmark corpus (Chelba et al., 2013) as our self-supervised English data-set. Text data in one Billion Word Language Model Benchmark corpus is obtained from the WMT11 website.[17] Normalization and tokenization were applied to the data, and duplicated sentences were removed. The vocabulary was constructed by discarding all words with a count below three, and sentence order was randomized. Finally, the corpus contains almost one billion words in the training data.

Table 2 shows the number of samples each dataset used in self-supervised learning. In total, we have around 40 million samples for this self-supervised training. One Billion Word Language Model Benchmark corpus has more than 30 million data samples and is the corpus with the most number of samples. Among the programming language datasets, CodeSearchNet Corpus is the most extensive corpus. When comparing only the programming languages, the Java language has the most self-supervised samples with more than two million inputs. Following them, Javascript and Python, where each have more than one million samples. Ruby, SQL, and LISP have the least number of self-supervised inputs. They only have around 150,000 samples or fewer, each.

---
14. https://github.com/LittleYUYU/StackOverflow-Question-Code-Dataset
15. https://github.com/topics/lisp?o=desc&s=stars
16. https://docs.github.com/en/rest/reference/repos#contents
17. http://statmt.org/wmt11/training-monolingual.tgz

### 2.3 Dataset Preprocessing

We mainly followed the instructions to preprocess the datasets if these guides existed with their original publications. We used the same parsers and tokenizers as the original repositories. We first removed the programs' comments for those without instructions or guidance because they are not part of the code. We applied different parsers to parse the code and get the code structure and element for programming language samples. Also, we substituted characters, strings, and numbers (integer, real, hexadecimal) with specific tokens in the code. Furthermore, we replaced the newline characters (\n, \r, \r\n) as $<newline>$. We also tokenized each sample and concatenated tokens in the sample to a sentence by inserting a space character between every two tokens to make all the samples from different datasets consistent at the input-level. Lastly, as an important point, we added a unique prefix for each task to allow models, specifically multi-task models, to distinguish between training samples from different tasks. For example, in code documentation generation for Javascript, we added "function documentation generation javascript: " as a prefix for all samples on this task.

- **English**: We used the *tokenize*[18] package from the NLTK Natural Language Toolkit (Bird, 2006). Most of the tasks' datasets contain at least one part English sentence, so this tokenizer is used in preprocessing these English texts. However, we did not apply it to the English corpus because it was already tokenized. We have used the English tokenizer for commit messages and API sequences because we do not have a specific tokenizer for them.

- **Python**: For the Python functions from CodeSearchNet Corpus Collection,[19] *tree-sitter* library for Python[20] was applied in order to have fair comparison between our results and theirs. For other Python codes, we used the Python *tokenize*[21] library that contains functions to separate the tokens of a string and returns the token's value and the type (number, string, etc.). For the rest of the tasks, we used it because it is the official library from Python, which is well maintained and updated regulary.

- **Java**: For the Java functions from CodeSearchNet Corpus Collection, *tree-sitter* library to parse Java[22] functions was applied. In addition, for other tasks involving the

---

18. https://www.nltk.org/api/nltk.tokenize.html
19. https://github.com/github/CodeSearchNet
20. https://github.com/tree-sitter/tree-sitter-python
21. https://docs.python.org/3.7/library/tokenize.html
22. https://github.com/tree-sitter/tree-sitter-java

Elnaggar, Ding, Matthes, Rost, Jones, Gibbs, Feher, Angerer, and SeveriniJava language, we used the Python library called *javalang*[23] that provides a lexer and a parser targeting Java.

- **Php, Go, Javascript, Ruby**: These languages are all involved in the CodeSearchNet Corpus Collection. They were parsed by Husain et al. (2019) using modified parsers based on the *tree-sitter* library for Php,[24] Go,[25] and Javascript.[26] We followed their methods and didn't make any changes.

- **SQL and CSharp**: The Python library *sqlparse*[27] was used for SQL tasks. This library provides support for parsing, splitting, and formatting SQL statements. For CSharp code, *ANTLR*[28] (ANother Tool for Language Recognition) parser from Parr (2013) was applied.

## 3. CodeTrans

We explain our models, the vocabulary generation steps, and the hardware we applied in this section.

### 3.1 Model

We adapted the encoder-decoder model proposed by Vaswani et al. (2017) and the T5[29] framework implemented by Raffel et al. (2020) to our tasks. The original T5 publication proposed five different model sizes — Small, Base, Large, 3B, and 11B. We used the small (60 million parameters), base (220 million parameters), and large model (770 million parameters) in this research. More details for the different models' parameters are shown in Table 3.

We set all the models' input and output length as 512 because most of the samples' length have less than 512 tokens. For the self-supervised objective, we applied the span-corruption strategy with a corruption rate of 15%. We considered a span of an average of three corrupted tokens as an entirety and used a unique mask token to replace it. For this model, the input consisted of the original input, but with some 3-gram words replaced by a unique mask token, while the target is the original corrupted 3-gram words surrounded by unique mask tokens for the uncorrupted spans. Different from the T5 models, we disabled the method of *reduce_concat_tokens*. By disabling this method, every sample will only have a single training example rather than concatenating different training examples up to the maximum training sequence length.

The T5 framework is very suitable for transfer learning, multi-task learning, and fine-tune the models. It has the concept of TaskRegistry and MixtureRegistry. Each task can be built as one TaskRegistry, and one or more TaskRegistries can build one MixtureRegistry. We built 13 TaskRegistries. Each programming language from each task has one

---

23. https://github.com/c2nes/javalang
24. https://github.com/tree-sitter/tree-sitter-php
25. https://github.com/tree-sitter/tree-sitter-go
26. https://github.com/tree-sitter/tree-sitter-javascript
27. https://github.com/andialbrecht/sqlparse
28. https://www.antlr.org/
29. https://github.com/google-research/text-to-text-transfer-transformer

|  |  | Small | Base | Large |
|---|---|---|---|---|
| Number of Blocks Each |  | 6 | 12 | 24 |
| Dense Layer Output Dimension |  | 2048 | 3072 | 4096 |
| Attention Layer Key Value Dimension |  | 64 | 64 | 64 |
| Number of Attention Heads |  | 8 | 12 | 16 |
| Sub-Layers and Embeddings Dimension |  | 512 | 768 | 1024 |
| Model Parameter (in Million) |  | 60 | 220 | 770 |
| Training Steps | Transfer Learning | 500,000 | 500,000 | 240,000 |
|  | Multi-Task Learning | 500,000 | 500,000 | 260,000 |
| Final Loss | Transfer Learning | 0.926 | 0.586 | 0.476 |
|  | Multi-Task Learning | 0.887 | 0.590 | 0.471 |
| Time Cost | Transfer Learning | 17 days | 53 days | 82 days |
|  | Multi-Task Learning | 17 days | 53 days | 87 days |

Table 3: Model Parameters for different size of models, as well as the time cost, and the final loss of transfer learning and multi-task learning pre-training stage with a batch size of 4,096.

TaskRegistry. We also built one MixtureRegistry for self-supervised learning and another MixtureRegistry for multi-task learning.

### 3.2 Vocabulary

Vocabulary is an essential aspect of natural language processing. Vocabulary itself contains much information about the corpus, like the corpus domain, formality, tone, and target audience. Vocabulary is helpful when processing the text corpus. We need to tokenize the input into different ids mapping to the components in the vocabulary before putting the text into the model. It is also the storage to construct the output of the model. The choice of vocabulary has a critical impact on model performance and output quality. Furthermore, the token frequency in the vocabulary also indicates the different importance of the text information.

We used the SentencePiece model (Kudo, 2018) to construct the vocabulary for this research, as well as to decode and encode the input/output. SentencePiece provides different tokenization methods, including the sub-word level tokenization. It extracts sub-words containing the semantic meanings and overcomes the drawback of the character level tokenization. The vocabulary generated could cover almost all the texts in the datasets. This is better than the word level tokenization, which requires an enormous vocabulary to cover most words in the datasets. We trained the SentencePiece on all the labeled and unlabeled datasets used in our experiment with the unigram language model algorithm. We set the id for padding token as 0, end of statement (EOS) token as 1, Unknown token as 2, and beginning of statement (BOS) token as 3. We set the size of the vocabulary to 32,000. The whole datasets have more than 46 million lines (each line could be considered one model input example and one SentencePiece input sentence). It is tremendous when using the unigram language model algorithm and would cause the training crash for training on the whole sentences. Therefore, we limited the "input sentence size" to 40 million, shuffled the input sentences to get random sentence inputs, and enabled the setting for training a huge corpus. We set the character coverage as 0.9999 because the corpus may contain non-English characters or meaningless symbols. In this way, we could exclude these noises from the vocabulary.

We noticed a significant amount of tokens of tokens indicating the programming languages and processes from the generated vocabulary, including "function," "String," "var," "import," etc. Furthermore, it covered most of the English vocabulary. This means our generated vocabulary is suitable for both natural language processing tasks and software engineering tasks.

### 3.3 Hardware

We utilized both Graphics Processing Units (GPUs) and Tensor Processing Units (TPUs) for training and evaluating the models. We used one NVIDIA GPU Quadro RTX 8000,[30] which has 576 NVIDIA Tensor Cores, 72 NVIDIA RT Cores, and 48 GB GDDR6 with ECC GPU memory. We used this GPU for all the single-task learning for small models and for part of the base models. We had two types of Google TPUs, v2-8, and v3-8. We obtained access to two TPUs v2-8 through the Google Colab notebooks[31] and multiple TPUs v3-8 using Google Cloud console. TPUs v3-8 are mainly used for multi-task learning, transfer learning pre-training, and fine-tuning models for large datasets. Moreover, TPUs v2-8 are applied for single-task training for the base model and fine-tuning the pre-trained models on relatively small datasets.

## 4. Experiments

We clarify our experiment details in this section, including single-task learning, transfer learning, multi-task learning, and multi-task learning with fine-tuning.

### 4.1 Single-Task Learning

For single-task learning, we trained the six tasks (13 sub-tasks in total) separately using the T5 framework. We trained both small and base size models, which generated two models for each task and, in total, 26 models. We tuned the batch size using the grid search inside the range of $2^5$ and $2^{10}$. We determined the training steps using early stopping concerning the models' performance on the validation sets based on the T5 built-in BLEU (Post, 2018) and ROUGE (Lin, 2004) scores. The optimal training steps with the corresponding batch size are listed in Table 4.

We noticed the following points during the single-task training:

- **The number of samples** in a data-set has an essential impact on the **model size** and the **training steps**. Task API Sequence Recommendation and Code Comment Generation have the two largest datasets. The small models for these two tasks require almost seven times more training steps than the base models until they could converge.

---

30. https://www.nvidia.com/en-us/design-visualization/quadro/rtx-8000/
31. https://colab.research.google.com/notebooks/intro.ipynb#recent=true

- Corpus for Source Code Summarization converges extremely fast. For SQL and CSharp data-sets in this corpus, the base model converges in 500 training steps even if the batch size is only 32, and the model has not seen the complete dataset yet. The scores on this task's validation set become worse if we train the model with more steps. So it is very **easy to overfit** the models for the Source Code Summarization task, mainly because of its small dataset size.

- Half of the models achieve the best performance with **a batch size of 256**. However, it varies slightly among different tasks, like the Source Code Summarization task requiring small batch sizes. Nevertheless, large batch sizes do not result in better performance, no matter the number of samples in the dataset.

### 4.2 Transfer Learning

Transfer Learning has two steps, pre-training and fine-tuning. The first uses the self-supervised method on unlabeled data, and the latter fine-tunes the models on supervised tasks using labeled datasets. We trained the small, base, and large T5 models for transfer learning.

All the self-supervised tasks were used and combined in the pre-training step. We set the T5 model to mask the spans of input data by enabling, which makes the model predict the masked content and builds an initial language model in this way. Since our pre-trained models used the datasets containing nine programming languages and one human language, these models are suitable to be fine-tuned on other downstream tasks in the software development domain and natural language processing domain. This can be seen as an expansion to current multi-language language models, which focuses on having a single model that supports only multiple human languages Xue et al. (2020). For per-training, we chose the batch size of 4096. We pre-trained the small and base model for 500,000 steps and the large model for 240,000 steps. Furthermore, we mainly used single TPU v3-8 in the pre-training. Table 3 shows the final loss and training time for different sizes of CodeTrans models during the pre-training.

After obtaining the pre-training model on the 500,000 training steps for the small and base models and 240,000 steps for the large model, we fine-tuned the models for the 13 supervised sub-tasks. We have noticed that half of the single-task learning models reach their best performance with a batch size of 256. So we chose 256 as the batch size for fine-tuning the downstream tasks. We applied early stopping to determine the fine-tuning steps based on the models' performance on the validation sets using the T5 built-in BLEU and ROUGE scores.

### 4.3 Multi-Task Learning

Multi-task learning trains a single model on a mixture of tasks, allowing sharing all model parameters across different tasks. This training strategy improves the data augmentation, focuses attention, and shares information for eavesdropping, learning the representation, and regularizes the weights (Ruder, 2017). Furthermore, it allows a single model to perform more than one task using the same weights. We trained 13 supervised sub-tasks together with all the self-supervised tasks. The self-supervised tasks are desired to help the model gain information about the language attributes and build a language model in both the software development domain and human language. This allows the model to understand both human written languages, in our case the English language, and the computer code languages. Simultaneously, the supervised tasks help each other make the model more generalized for all the tasks and avoid overfitting on each specific task.

| Task | Language | Sample Size | Model Size | Batch Size | | | Training Steps | | |
|---|---|---|---|---|---|---|---|---|---|
| | | | | ST | TF-FT | MT-FT | ST | TF-FT | MT-FT |
| Code Documentation Generation | Python | 251,820 | Small | 256 | 256 | 256 | 20,000 | 5,000 | 4,000 |
| | | | Base | 384 | 256 | 256 | 90,000 | 2,000 | 4,000 |
| | | | Large | - | 256 | 256 | - | 500 | 500 |
| | Java | 164,923 | Small | 256 | 256 | 256 | 60,000 | 10,000 | 2,000 |
| | | | Base | 256 | 256 | 256 | 80,000 | 5,000 | 2,000 |
| | | | Large | - | 256 | 256 | - | 500 | 500 |
| | Go | 167,288 | Small | 256 | 256 | 256 | 5,000 | 10,000 | 2,000 |
| | | | Base | 256 | 256 | 256 | 80,000 | 5,000 | 2,000 |
| | | | Large | - | 256 | 256 | - | 1,000 | 4,500 |
| | Php | 241,241 | Small | 256 | 256 | 256 | 200,000 | 10,000 | 2,000 |
| | | | Base | 1024 | 256 | 256 | 30,000 | 65,000 | 5,000 |
| | | | Large | - | 256 | 256 | - | 18,000 | 8,000 |
| | Ruby | 24,927 | Small | 128 | 256 | 256 | 10,000 | 5,000 | 2,000 |
| | | | Base | 128 | 256 | 256 | 8,000 | 5,000 | 12,000 |
| | | | Large | - | 256 | 256 | - | 1,000 | 2,000 |
| | Javascript | 58,023 | Small | 256 | 256 | 256 | 16,000 | 40,000 | 32,000 |
| | | | Base | 256 | 256 | 256 | 18,000 | 35,000 | 10,000 |
| | | | Large | - | 256 | 256 | - | 4,000 | 2,500 |
| Source Code Summarization | Python | 12,004 | Small | 233 | 256 | 256 | 5,000 | 5,000 | 600 |
| | | | Base | 32 | 256 | 256 | 1,000 | 1,000 | 1,000 |
| | | | Large | - | 256 | 256 | - | 100 | 100 |
| | Csharp | 52,943 | Small | 128 | 256 | 256 | 2,000 | 2,000 | 1,200 |
| | | | Base | 32 | 256 | 256 | 500 | 500 | 500 |
| | | | Large | - | 256 | 256 | - | 200 | 100 |
| | SQL | 25,671 | Small | 128 | 256 | 256 | 500 | 1,000 | 1,200 |
| | | | Base | 32 | 256 | 256 | 500 | 500 | 500 |
| | | | Large | - | 256 | 256 | - | 200 | 100 |
| Code Comment Generation | Java | 470,451 | Small | 256 | 256 | 256 | 520,000 | 750,000 | 750,000 |
| | | | Base | 256 | 256 | 256 | 80,000 | 80,000 | 60,000 |
| | | | Large | - | 256 | 256 | - | 60,000 | 25,000 |
| Git Commit Message Generation | Java | 26,208 | Small | 128 | 256 | 256 | 15,000 | 5,000 | 8,000 |
| | | | Base | 512 | 256 | 256 | 4,000 | 2,000 | 16,000 |
| | | | Large | - | 256 | 256 | - | 4,500 | 3,000 |
| API Sequence Recommendation | Java | 7,475,850 | Small | 256 | 256 | 256 | 840,000 | 1,400,000 | 1,150,000 |
| | | | Base | 256 | 256 | 256 | 145,000 | 340,000 | 320,000 |
| | | | Large | - | 256 | 256 | - | 180,000 | 130,000 |
| Program Synthesis | DSL | 79,214 | Small | 512 | 256 | 256 | 6,000 | 5,000 | 16,000 |
| | | | Base | 256 | 256 | 256 | 10,000 | 45,000 | 30,000 |
| | | | Large | - | 256 | 256 | - | 3,500 | 2000 |

Table 4: The single-task learning (ST) training steps, transfer learning fine-tuning (TF-FT) steps, and multi-task learning fine-tuning (MT-FT) steps for the small, base, and large models. We also listed the batch size of each task in this table. The batch sizes of single-task learning are different for different tasks. However, the batch size of transfer learning and multi-task learning are both 256.

We used examples-proportional mixing to select samples in proportion to the size of each task's dataset and concatenated them. This ensures that the model will see samples from small datasets as it will see samples from large datasets on every batch. We recorded the model checkpoint every 20,000 training steps, using a batch size of 4,096. Usually, all the tasks should share one same best performance checkpoint. However, Raffel et al. (2020) proposed a way to relax this goal and select a different checkpoint for each task. We also evaluated the model on the validation set and selected the best checkpoint for each task. Therefore, each task could have a different checkpoint from the same model. We trained T5 small, base, and large models using only single TPU v3-8. Table 3 illustrate the training steps, final loss, and the time cost for the multi-task learning models.

### 4.4 Multi-Task Learning with Fine-Tuning

We further fine-tuned the multi-task learning final checkpoint, 500k steps for the small and base model, and 260k steps for the large model, for each supervised task separately. Like the transfer learning fine-tuning, we chose the batch size of 256 and applied early stopping to determine the fine-tuning steps based on the models' performance on the validation datasets. Table 4 shows all the related hyper-parameters for the multi-task learning fine-tuning step.

## 5. Result

We ran the final evaluation on the test dataset, and we compared the performance of CodeTrans models with different state-of-the-art models for each task. Therefore, we applied the same metrics as other SOT models for calculating the evaluation results, as shown in Table 5.

### 5.1 Code Documentation Generation

We evaluated the Code Documentation Generation tasks using smoothed BLEU-4 score (Lin and Och, 2004) on the CodeSearchNet test dataset. The evaluation results are shown in Table 5, and were compared with CodeBert (Feng et al., 2020). Overall, we outperformed CodeBert on all the programming languages in this task. Multi-task learning has, in general, the best performance and achieves the best result for three programming languages. The possible reason for this is that the CodeSearchNet dataset is involved in two self-supervised tasks during the multi-task training, and it has seen many training data related to this task. Transfer learning and multi-task learning fine-tuning CodeTrans models also have relatively good performance and are much better than single-task learning.

### 5.2 Source Code Summarization

We also applied smoothed BLEU-4 to evaluate the Source Code Summarization task. We compared CodeTrans performance with Code-NN (Iyer et al., 2016). Unfortunately, Code-NN did not provide their result for the Python code; however, we did evaluate it for future comparisons. By evaluating the SQL and CSharp code, Code-NN selected 100 samples with two additional human annotations and calculated the smoothed BLEU-4 on these samples. We followed their instruction and evaluated in the same way to have a fair comparison. CodeTrans outperformed Code-NN on the existing scores as shown in Table 5. Among different CodeTrans models, multi-task learning has the best performance on two of the programming languages for this task.

### 5.3 Code Comment Generation

We compared CodeTrans' performance with DeepCom using smoothed BLEU-4 for the task Code Comment generation, with the results shown in Table 5. CodeTrans transfer learning large model has the best performance, and its smoothed BLEU score is higher than DeepCom by more than one percent. CodeTrans models with multi-task learning have the worst performance. However, the score increases with the increase of model size. The reason is the Code Comment Generation dataset has the second-largest sample size with 470,451 samples, and we need bigger models to exploit them all.

| | Code Documentation Generation | | | | | | Source Code Summarization | | | | CCG | Git-Gen | API-Gen | PS |
|---|---|---|---|---|---|---|---|---|---|---|---|---|---|---|
| | Python | Java | Go | Php | Ruby | Javascript | Python | SQL | CSharp | | Java | Java | Java | DSL |
| CodeTrans-ST-Small | 17.31 | 16.65 | 16.89 | 23.05 | 9.19 | 13.70 | 8.45 | 17.55 | 19.74 | 37.98 | 39.61 | 68.71 | 89.43 |
| CodeTrans-ST-Base | 16.86 | 17.17 | 17.16 | 22.98 | 8.23 | 13.17 | 9.12 | 15.00 | 18.65 | 38.07 | 38.67 | 70.45 | 89.65 |
| CodeTrans-TF-Small | 19.93 | 19.48 | 18.88 | 25.35 | 13.15 | 17.23 | 10.06 | 17.71 | 20.40 | 38.56 | 44.22 | 68.90 | 90.30 |
| CodeTrans-TF-Base | 20.26 | 20.19 | 19.50 | 25.84 | 14.07 | 18.25 | 10.94 | 17.66 | 21.12 | 39.06 | 44.17 | 72.11 | 90.24 |
| CodeTrans-TF-Large | 20.35 | 20.06 | **19.54** | 26.18 | 14.94 | **18.98** | 12.41 | 18.40 | 21.43 | **39.50** | **44.41** | 73.26 | 90.21 |
| CodeTrans-MT-Small | 19.64 | 19.00 | 19.15 | 24.68 | 14.91 | 15.26 | 13.11 | 19.15 | 22.39 | 20.15 | 36.17 | 58.43 | 82.88 |
| CodeTrans-MT-Base | **20.39** | 21.22 | 19.43 | **26.23** | **15.26** | 16.11 | **13.37** | 19.24 | **23.20** | 27.44 | 39.25 | 67.97 | 86.99 |
| CodeTrans-MT-Large | 20.18 | **21.87** | 19.38 | 26.08 | 15.00 | 16.23 | 13.24 | 19.49 | 23.57 | 34.69 | 41.18 | 72.29 | 90.27 |
| CodeTrans-MT-TF-Small | 19.77 | 20.04 | 19.36 | 25.55 | 13.70 | 17.24 | 12.10 | 18.25 | 22.03 | 38.37 | 43.96 | 69.29 | **90.31** |
| CodeTrans-MT-TF-Base | 19.77 | 21.12 | 18.86 | 25.79 | 14.24 | 18.62 | 10.64 | 16.91 | 21.40 | 38.90 | 44.19 | 72.89 | 90.30 |
| CodeTrans-MT-TF-Large | 18.94 | 21.42 | 18.77 | 26.20 | 14.19 | 18.83 | 12.14 | **19.98** | 21.10 | 39.25 | 44.34 | **73.39** | 90.17 |
| CodeBert(Feng et al., 2020) | 19.06 | 17.65 | 18.07 | 25.16 | 12.16 | 14.90 | - | - | - | - | - | - | - |
| CODE-NN(Iyer et al., 2016) | - | - | - | - | - | - | - | 18.40 | 20.50 | - | - | - | - |
| DeepCom(Hu et al., 2018) | - | - | - | - | - | - | - | - | - | 38.17 | - | - | - |
| NMT(Jiang et al., 2017) | - | - | - | - | - | - | - | - | - | - | 32.81 | - | - |
| DeepAPI(Gu et al., 2016) | - | - | - | - | - | - | - | - | - | - | - | 54.42 | - |
| Seq2Tree(Polosukhin and Skidanov, 2018) | - | - | - | - | - | - | - | - | - | - | - | - | 85.80 |

Table 5: Evaluation results of all the tasks in this paper. CCG means the task Code Comment Generation. Git-Gen is Git Commit Message Generation. API-Gen stands for the task API Sequence Recommendation. PS is the task Program Synthesis. ST stands for single-task learning, TF is transfer learning, MT means multi-task learning, and MT-FT is multi-task learning with fine-tuning. We evaluate the CodeTrans on the test dataset and compare the performance of CodeTrans with state-of-the-art models. We applied smoothed BLEU-4 for Code Documentation Generation, Source Code Summarization, and Code Comment Generation. We use BLEU-4 for Git Commit Message Generation and API Sequence Recommendation, and accuracy for Program Synthesis.

### 5.4 Git Commit Message Generation

The task Git Commit Message Generation's evaluation result is listed in Table 5. We applied the BLEU-4 (Post, 2018) for the evaluation to compare it with previous research results. All the CodeTrans models, including single-task training, outperform the NMT model. Among them, the CodeTrans transfer learning large model has the best BLEU-4 score. The performance of the CodeTrans multi-task learning large model is very close to the transfer learning large model.

### 5.5 API Sequence Recommendation

Table 5 also contains the task API Sequence Recommendation evaluation result. We compared the CodeTrans models with the DeepAPI model. We applied the same BLEU-4 metric script as the DeepAPI used. All the CodeTrans models, including single-task training, outperform the DeepAPI model. Among the CodeTrans models, those trained using only multi-task learning performed the most poorly mainly because of this task's large data-set. However, with increasing multi-task learning model size, it started to reach a close/better performance than single-task learning and transfer learning. The CodeTrans large model with multi-task learning fine-tuning has the highest scores across all the models. The CodeTrans transfer learning large model also has a similarly good performance.

### 5.6 Program Synthesis

We used accuracy when evaluating the Program Synthesis task. This accuracy calculates whether our model output is precisely the same as the golden reference. Seq2Tree used the code accuracy to count how many of the model's outputs can pass the code tests. If the model output is identical to the reference, this output could pass the code tests. As shown in Table 5, nine out of ten CodeTrans models outperform the Seq2Tree model. The CodeTrans multi-task learning fine-tuning small model achieves the best score on accuracy. For multi-task learning, the performance increases along with the model size. However, for transfer learning and multi-task fine-tuning, smaller models perform better. Generally, this task's scores are very high, which means that this is an easy task with similar validation and test sets, and bigger models may be easy to be overfitted.

## 6. Reflection

We discuss our results and provides a conclusion in this section. We also list our future work here.

### 6.1 Discussion

Our CodeTrans models with the transformer architecture with both encoder and decoder outperform the thirteen sub-tasks' baseline models. This proves the effectiveness of the transformer encoder-decoder architecture for these tasks in the software engineering domain, mainly when we utilize transfer learning and multi-task learning methods with self-supervised learning. Nevertheless, the models' performance varies a bit when using different training strategies for different sizes of models on different datasets.

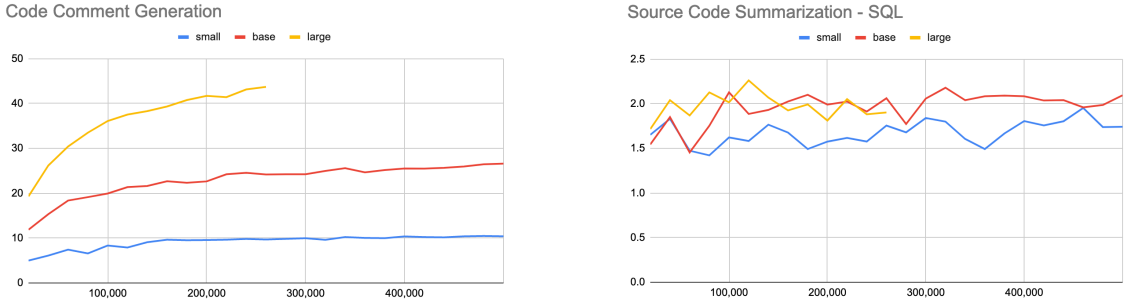

(a) The Code Comment Generation task's training dataset has 470,486 samples.

(b) The Source Code Summarization - SQL task's training dataset has 22,492 samples.

Figure 1: The evaluation of multi-task learning checkpoints on the validation set for two tasks. The x-axis lists the training steps. The y-axis is the T5 built-in BLEU score. Different colors indicate different sizes of models.

We noticed that the **model size** plays an essential role in the model's performance. For single-task learning, the larger the dataset is, the fewer training steps a bigger model requires. A bigger model reaches a lower loss under the same batch size and the same evaluation steps when applying the multi-task learning or transfer learning strategy. Although the pre-training may cost more time for bigger models, they need fewer iteration steps during fine-tuning for each task than the small models. As a result, for most of the tasks, the bigger the model is, the better the evaluation scores the model could achieve with even less fine-tuning time.

The evaluation results also prove that **transfer learning** and **multi-task learning with fine-tuning** strategies outperform the models that only used **single-task learning** on all the tasks. The performance of models using transfer learning is very similar to those using multi-task learning fine-tuning. It is hard to say which one is better. However, transfer learning does not require the task dataset to be involved in the pre-training steps. For a new task, the dataset only needs to be trained for relatively few fine-tuning steps, while multi-task learning with fine-tuning needs the new task dataset during pre-training. We can clearly say that transfer learning would save time and provide better results for a new task when fine-tuned on a pre-trained model checkpoint.

The performance of **multi-task learning** depends heavily on the **data size** and **attributes** of the task itself. For **large datasets** like the datasets for the task Code Comment Generation and API Sequence Recommendation, multi-task learning models are even worse than the models that only applied single-task learning as shown in Table 5. Figure 1a shows that the model's performance improves significantly when we increase the model size for the Code Comment Generation task with a large dataset. Half a million multi-task training steps are not enough for this task, even using the large model. When the dataset is **tiny and easy to be overfitted**, small multi-task models could achieve the best result while bigger setups do not always lead to better performance. This is due to the higher probability of overfitting the small dataset when more parameters are used, even when we used regularization methods like a dropout (10% in all our models). Figure 1b shows that the base model performs better overall than the small models for the source code summarization - SQL task, but the large model has several overlaps with the base model. The large model has a sign of overfitting after 120,000 training steps, and the model's performance decreases after that point.

| Model | Size | Model Output |
| --- | --- | --- |
| CodeTrans Single-Task Learning | Small | mysql : how to get the difference of a column in a table ? |
| | Base | how do i get the average of a date range in sql server 2005 ? |
| CodeTrans Transfer Learning | Small | how to get the time in milliseconds since the start time of the transaction was taken ? |
| | Base | how to get current date time in sql server ? |
| | Large | mysql time ( ) function |
| CodeTrans Multi-Task Learning | Small | how to get the time in mysql ? |
| | Base | how can i get the time of a date in mysql ? |
| | Large | how to convert datetime to time in mysql ? |
| CodeTrans Multi-Task Learning Fine-tuning | Small | how to get the correct time from mysql database ? |
| | Base | how to convert date to time in mysql ? |
| | Large | select time from mysql table |
| Code Snippet as Input | | select time ( col0 ) from tab0 |
| Golden Reference | | datetime implementation in php mysql |

Table 6: The models' output for an example of the task Source Code Summarization. We compared different CodeTrans model outputs and the golden reference for the input SQL code "select time ( col0 ) from tab0". The golden reference is the one extracted from the StackOverflow.

Table 6 lists each CodeTrans model's outputs compared with the golden reference extracted from the StackOverflow. The input for the models is "select time ( col0 ) from tab0". We can observe that all the models' outputs are readable sentences. The majority of them have a question format. Because the dataset contains questions and answers from StackOverflow, the models have learned how to ask questions. Outputs from single-task learning models do not make much sense. The transfer learning and multi-task learning outputs all notice that this code is about *time*. All the multi-task learning models also specified the *mysql* database system. The CodeTrans multi-task learning large model mentions the keyword *datetime*, which also appears in the golden reference. Besides, the transfer learning and multi-task learning fine-tuning base models have reasonable outputs as well. The CodeTrans transfer learning and multi-task learning fine-tuning models focus more on the code function and structure to summarize this code snippet. In total, our judgement for the models' performances matches the ranking of our evaluation metrics. For more examples, we chosen one example from each task and list all the models' output for this example in the Appendix.

Moreover, most of the Code Documentation Generation tasks achieved the best evaluation performance when using the multi-task learning strategy. It is possible that we have two more self-supervised tasks from the same CodeSearchNet corpus during the multi-task learning. These give more similar samples for the supervised Code Documentation Generation tasks so that the model would focus on performing better for these tasks. Furthermore, using different types of tasks during multi-task learning efficiently avoids overfitting.

### 6.2 Conclusion

This paper explores the CodeTrans models with transformer encoder-decoder architecture on six main tasks and, in total, thirteen sub-tasks in the software engineering domain covering nine programming languages. We have carried out experiments with different training strategies, including single-task learning, transfer learning, multi-task learning, and multi-task learning with fine-tuning. We applied different models' sizes based on the Text-To-Text Transfer Transformer framework by utilizing NVIDIA GPUs and Google Cloud TPUs.

Our CodeTrans models outperform all the baseline models and achieve the state of the art over all the tasks. Our experiments on various tasks provide us with many insights about training a neural network model on software engineering-relevant tasks. First, we find that larger models can bring a better model performance. Second, models with transfer learning perform as well as models with multi-task learning fine-tuning, and the pre-training models can be fine-tuned on the new downstream tasks efficiently while saving a significant amount of training time. Moreover, multi-task learning is very beneficial for the small dataset on which the model will overfit easily. It is very promising that these experiences can be generalized for training natural language processing tasks on different domains.

In addition to these findings, we published our models on the Hugging Face Hub so that everyone can access our models and use them for their purposes. We also provided online the pre-trained checkpoints generated from our CodeTrans during transfer learning pre-training and multi-task learning. These checkpoints are suitable for fine-tuning other software engineering tasks if the task's programming language is covered in this paper.

### 6.3 Future Work

When working on the Code Documentation Generation tasks, we have noticed that a programming language function has two aspects influencing the model performance: the function names/parameter names and the code structure. A well-named function would lower the difficulty for the model to generate the documentation. Further investigation about functions with disguised parameter names or function names would be valuable. In this work, we considered a function as a sentence; from this aspect, we do not fully make use of the code structure, so how to present the code is also a good research point. Experiments about finding the best way to present the features of code structure can be carried out.

We preprocessed the datasets by parsing and tokenizing the programming codes using different Python libraries for each programming language. So when using our models, applying the same preprocessing way would draw the best results. Nevertheless, not every user is a programming expert, and the preprocessing increases the complexity for users to get the best model performance. It would be meaningful to examine the effect of preprocessing for the software engineering tasks and train models with good performance, but without preprocessing like parsing and tokenizing.

Moreover, more tasks can be explored using transformer encoder-decoder architecture. It would be interesting to examine our models' performance on the unseen programming languages as Feng et al. (2020) did in their experiments. Furthermore, testing the pre-trained language models on human language tasks like SQuAD dataset Rajpurkar et al.

(2016) could be of interest. Evaluation could be run directly on similar tasks with an unseen programming language using the multi-task learning CodeTrans models.

**Acknowledgments**

The authors primarily thank Tim Karl (TUM) and Jian Kong (TUM) for their invaluable help with hardware and software; Inga Weise and Aline Schmidt (both TUM) for support with many other aspects of this work. Thanks for invaluable support and feedback from NVIDIA, in particular to Ulrich Michaelis, Ada Sedova, Geetika Gupta, Axel Koehler, Frederic Pariente, Jonathan Lefman, and Thomas Bradley. From Google, we would like to deeply thank Jamie Kinney, Alex Schroeder, Nicole DeSantis, Andrew Stein, Vishal Mishra, Eleazar Ortiz, Nora Limbourg, Cristian Mezzanotte, and all TFRC Team for their invaluable support to set up our projects on Google Cloud and solve all the related Google TPU and servers issues. All CodeTrans models could not be easily publicly accessible without the amazing support from the Hugging Face team; that is why we are very grateful to Patrick von Platen, Julien Chaumond, and Clément Delangue from Hugging Face. Last, but not least, thanks to all those who deposit their experimental data in public databases, and to those who maintain these databases.

This work was supported by a grant from Software Campus through the German Ministry for Research and Education (BMBF: Bundesministerium fuer Bildung und Forschung). We gratefully acknowledge the support of NVIDIA Corporation with the donation of two Titan GPU used for this research development phase. We also want to thank LRZ (Leibniz Rechenzentrum) for providing us access to DGX-1(V100) for the testing phase.

Finally and most importantly, this research used resources of TPUs under TensorFlow Research Cloud grant. Furthermore, Rostlab acknowledge support from Google Cloud and Google Cloud Research Credits program to fund this project under Covid19 grants.

# Appendix Model Output Examples for each Task.

For each sub-task we list one example containing all the models' output, the input, and the standard reference in this Appendix. These examples are all randomly selected from the supervised tasks test datasets.

## Appendix A. Code Documentation Generation

### A.1 Code Documentation Generation — Python

| Model | Size | Model Output |
|---|---|---|
| CodeTrans Single-Task Learning | Small | Prints an error and exits with an optional exit code . |
| | Base | Prints the e . |
| CodeTrans Transfer Learning | Small | Prints a message and exits with the specified exit code . |
| | Base | Prints an error message and exits . |
| | Large | Print an error message and exit . |
| CodeTrans Multi-Task Learning | Small | Prints a message and exits if exit_code is not None . |
| | Base | Prints an error message and exits with the given exit code . |
| | Large | Print an error message and exit . |
| CodeTrans Multi-Task Learning Fine-Tuning | Small | Print a message and exit . |
| | Base | Prints a message to the log and exits with an optional exit code . |
| | Large | Print error message and exit . |
| Code Snippet as Input | | def e ( message , exit_code = None ) : print_log ( message , YELLOW , BOLD ) if exit_code is not None : sys . exit ( exit_code ) |
| Golden Reference | | Print an error log message . |

### A.2 Code Documentation Generation — Java

| Model | Size | Model Output |
|---|---|---|
| CodeTrans Single-Task Learning | Small | Casts a Cast to another type |
| | Base | Casts a value into a function that returns the same type as the supplied parameter . |
| CodeTrans Transfer Learning | Small | Casts the given class to a function . |
| | Base | Casts a function to a class |
| | Large | Cast a value to the given target type . |
| CodeTrans Multi-Task Learning | Small | Casts the given class to the given target class . |
| | Base | Returns a function that casts the given value to the given type . |
| | Large | Creates a function that performs a cast to the given class . |
| CodeTrans Multi-Task Learning Fine-Tuning | Small | Creates a function that casts the given object to the target type . |
| | Base | Returns a function that casts the given object to the given class . |
| | Large | Returns a function that casts its input type to the target type . |
| Code Snippet as Input | | public static ¡ T , U ¿ Function ¡ T , U ¿ castFunction ( Class ¡ U ¿ target ) { return new CastToClass ¡ T , U ¿ ( target ) ; } |
| Golden Reference | | Returns a function that cast the incoming values via a Class object . |

## A.3 Code Documentation Generation — Go

| Model | Size | Model Output |
| --- | --- | --- |
| CodeTrans Single-Task Learning | Small | needSnapshotAbort returns true if we need to roll a snapshot . |
| | Base | needSnapshotAbort returns true if we need to call snapshot and false otherwise . |
| CodeTrans Transfer Learning | Small | needSnapshotAbort returns true if the Progress . Match field of ProgressState == ProgressStateSnapshot and pr . Match ¡ pr . PendingSnapshot . |
| | Base | needSnapshotAbort returns true if the progress is in a snapshot and the match exceeds the pending snapshot . |
| | Large | needsnapshotabort returns true if we need to abort a snapshot . |
| CodeTrans Multi-Task Learning | Small | needSnapshotAbort returns true if the current state is a snapshot . |
| | Base | needSnapshotAbort returns true if we need to abort the snapshot . |
| | Large | needsnapshotabort returns true if we need to abort a snapshot . |
| CodeTrans Multi-Task Learning Fine-Tuning | Small | needSnapshotAbort returns true if the progress needs to be aborted . |
| | Base | needSnapshotAbort returns true if we need to abort the snapshot |
| | Large | needsnapshotabort returns true if we need to abort the snapshot . |
| Code Snippet as Input | | func ( pr * Progress ) needSnapshotAbort ( ) bool { return pr . State == ProgressStateSnapshot && pr . Match ¿= pr . PendingSnapshot } |
| Golden Reference | | needSnapshotAbort returns true if snapshot progress s Match is equal or higher than the pendingSnapshot . |

## A.4 Code Documentation Generation — Php

| Model | Size | Model Output |
| --- | --- | --- |
| CodeTrans Single-Task Learning | Small | Update table with specific data |
| | Base | Call when table is constructed |
| CodeTrans Transfer Learning | Small | Updates a table . |
| | Base | Calls the updateTable method to update a table . |
| | Large | Updates a table . |
| CodeTrans Multi-Task Learning | Small | Update a table . |
| | Base | Updates a table . |
| | Large | updates a database table . |
| CodeTrans Multi-Task Learning Fine-Tuning | Small | Update a table . |
| | Base | Updates a table . |
| | Large | updates a database table . |
| Code Snippet as Input | | public static function update ( $ table ) { if ( ! is_array ( $ table ) ) { $ table = json_decode ( $ table , true ) ; } if ( ! SchemaManager :: tableExists ( $ table [ 'oldName' ] ) ) { throw SchemaException :: tableDoesNotExist ( $ table [ 'oldName' ] ) ; } $ updater = new self ( $ table ) ; $ updater -¿ updateTable ( ) ; } } |
| Golden Reference | | Update the table . |

## A.5 Code Documentation Generation — Ruby

| Model | Size | Model Output |
| --- | --- | --- |
| CodeTrans Single-Task Learning | Small | Log a message at the given level if the logger is present |
| | Base | Writes a log message if the current log level is at or below the supplied severity . |
| CodeTrans Transfer Learning | Small | Create an IO object at the given level with the given block . This method will not be called directly if the IO object at that level . |
| | Base | Writes a message of the given severity to the log . |
| | Large | Writes a message to the log device . |
| CodeTrans Multi-Task Learning | Small | Add a message to the log . |
| | Base | Add a message to the logger . |
| | Large | Add a message to the log . |
| CodeTrans Multi-Task Learning Fine-Tuning | Small | Writes a message to the log with a given severity . |
| | Base | We need to overload the add method . Basibally it is the same as the original one but we add our own log format to it . |
| | Large | Writes a message if the severity is high enough . This method is executed asynchronously . |
| Code Snippet as Input | | def add ( severity , progname , & block ) return true if io . nil? —— severity ¡ level message = format_message ( severity , progname , yield ) MUTEX . synchronize { io . write ( message ) } true end |
| Golden Reference | | Add a message to the logger with the given severity and progname . |

## A.6 Code Documentation Generation — Javascript

| Model | Size | Model Output |
|---|---|---|
| CodeTrans Single-Task Learning | Small | Returns true if the browser is a native element . |
| | Base | Returns whether the givenEnv should be focused . |
| CodeTrans Transfer Learning | Small | Checks if the current browser is on a standard browser environment . |
| | Base | Check if browser environment is a standard browser environment |
| | Large | Check if the environment is standard browser . |
| CodeTrans Multi-Task Learning | Small | Returns true if the browser environment is a standard browser environment . |
| | Base | Checks if the current browser environment is a standard browser environment . |
| | Large | Determines if the current environment is a standard browser environment |
| CodeTrans Multi-Task Learning Fine-Tuning | Small | Standard browser environment has a notion of what React Native does not support it . |
| | Base | Check if the browserEnv is standard . |
| | Large | Checks if the browser is in a standard environment . |
| Code Snippet as Input | | function isStandardBrowserEnv ( ) { if ( typeof navigator !== 'undefined' && ( navigator . product === 'ReactNative' —— navigator . product === 'NativeScript' —— navigator . product === 'NS' ) ) { return false ; } return ( typeof window !== 'undefined' && typeof document !== 'undefined' ) ; } |
| Golden Reference | | Determine if we are running in a standard browser environment |

# Appendix B. Source Code Summarization

## B.1 Source Code Summarization — Python

| Model | Size | Model Output |
|---|---|---|
| CodeTrans Single-Task Learning | Small | Python : Compare files containing a folder |
| | Base | How to copy a text file into a variable ? |
| CodeTrans Transfer Learning | Small | includes within a line ' with python 2.7 |
| | Base | Python : Include specific lines in a file |
| | Large | Changing a line in a text file |
| CodeTrans Multi-Task Learning | Small | How to write a string to a file in Python ? |
| | Base | How to write a text file to a text file in Python ? |
| | Large | How to remove a line from a file in python ? |
| CodeTrans Multi-Task Learning Fine-Tuning | Small | How to include text in a file using Python ? |
| | Base | Python : 'Include this text in a file ' |
| | Large | How to remove empty string in python |
| Code Snippet as Input | | with open ( CODE_STRING , CODE_STRING ) as in_file : buf = in_file . readlines ( ) with open ( CODE_STRING , CODE_STRING ) as out_file : for line in buf : if line == " ; Include this text " : line = line + " Include below " out_file . write ( line ) ' |
| Golden Reference | | Adding lines after specific line |

## B.2 Source Code Summarization — CSharp

| Model | Size | Model Output |
|---|---|---|
| CodeTrans Single-Task Learning | Small | How to convert C # DateTime to C # |
| | Base | How to convert DateTime to DateTime format in C # ? |
| CodeTrans Transfer Learning | Small | How to convert double to DateTime in C # ? |
| | Base | Convert double to DateTime |
| | Large | How to convert double to datetime ? |
| CodeTrans Multi-Task Learning | Small | How to convert unix timestamp to local time in C # ? |
| | Base | How to convert a double to a DateTime in C # ? |
| | Large | How to convert double to datetime in C # ? |
| CodeTrans Multi-Task Learning Fine-Tuning | Small | How to convert double to DateTime in C # ? |
| | Base | How to convert double to DateTime in C # ? |
| | Large | Convert double to datetime in C # |
| Code Snippet as Input | | public static DateTime ParseUnixDateTime ( double unixTime ) { var dt = new DateTime ( CODE_INTEGER , CODE_INTEGER , CODE_INTEGER , CODE_INTEGER , CODE_INTEGER , CODE_INTEGER , CODE_INTEGER , System . DateTimeKind . Utc ) ; dt = dt . AddSeconds ( unixTimeStamp ) . ToLocalTime ( ) ; return dt ; } |
| Golden Reference | | C # Display compare timestamp to current month |

## B.3 Source Code Summarization — SQL

| Model | Size | Model Output |
|---|---|---|
| CodeTrans Single-Task Learning | Small | mysql : how to get the difference of a column in a table ? |
| | Base | how do i get the average of a date range in sql server 2005 ? |
| CodeTrans Transfer Learning | Small | how to get the time in milliseconds since the start time of the transaction was taken ? |
| | Base | how to get current date time in sql server ? |
| | Large | mysql time ( ) function |
| CodeTrans Multi-Task Learning | Small | how to get the time in mysql ? |
| | Base | how can i get the time of a date in mysql ? |
| | Large | how to convert datetime to time in mysql ? |
| CodeTrans Multi-Task Learning Fine-Tuning | Small | how to get the correct time from mysql database ? |
| | Base | how to convert date to time in mysql ? |
| | Large | select time from mysql table |
| Code Snippet as Input | | select time ( col0 ) from tab0 |
| Golden Reference | | datetime implementation in php mysql |

# Appendix C. Code Comment Generation

| Model | Size | Model Output |
|---|---|---|
| CodeTrans Single-Task Learning | Small | Renders the URI into its textual representation . |
| | Base | Renders the URI . |
| CodeTrans Transfer Learning | Small | Renders the given URI into a string . Override this if you want to do something special to convert the URI into ASCII . |
| | Base | Render a URI |
| | Large | Renders the given URI using the encoding scheme in ISO-8859-1 . |
| CodeTrans Multi-Task Learning | Small | Renders a URI . |
| | Base | Renders the given URI into a string . |
| | Large | Renders the URI as an ASCII string . |
| CodeTrans Multi-Task Learning Fine-Tuning | Small | Render the URI as ASCII string . |
| | Base | Renders a URI as a string , escaping special characters . |
| | Large | Render the given URI . |
| Code Snippet as Input | | protected String renderUri ( URI uri ) return uri . toASCIIString ( ) ; |
| Golden Reference | | Render the URI as a string |

# Appendix D. Git Commit Message Generation

| Model | Size | Model Output |
|---|---|---|
| CodeTrans Single-Task Learning | Small | Added missing Jungtaeks to log4j . jar |
| | Base | Added missing jar file |
| CodeTrans Transfer Learning | Small | Adding missing joscar jar |
| | Base | Added joscar . jar to the lib directory of the Gateway project |
| | Large | added missing joscar jar |
| CodeTrans Multi-Task Learning | Small | add joscar jar |
| | Base | Add joscar jar |
| | Large | Add joscar |
| CodeTrans Multi-Task Learning Fine-Tuning | Small | Adding missing jar file |
| | Base | Added joscar jar |
| | Large | Started work on the icq / aim protocol provider implementation |
| Code Snippet as Input | | new file mode 100644 index 000000000 . . 892fda21b Binary files / dev / null and b / src / plugins / gateway / lib / joscar . jar differ |
| Golden Reference | | Added joscar JAR |

CodeTrans: Transformer for Source Code## Appendix E. API Sequence Generation

| Model | Size | Model Output |
|---|---|---|
| CodeTrans Single-Task Learning | Small | Color . getRed Color . getGreen Color . getBlue |
| | Base | Color . getRed Color . getGreen Color . getBlue |
| CodeTrans Transfer Learning | Small | Color . getRed Color . getGreen Color . getBlue |
| | Base | Math . max Math . min |
| | Large | Math . max Math . min |
| CodeTrans Multi-Task Learning | Small | Math . min Math . max |
| | Base | Color . getRed Color . getGreen Color . getBlue |
| | Large | Renders Math . round Math . max Math . round Math . max Math . round |
| CodeTrans Multi-Task Learning Fine-Tuning | Small | Color . getRed Color . getGreen Color . getBlue |
| | Base | Color . getRed Color . getGreen Color . getBlue |
| | Large | Color . getRed Color . getGreen Color . getBlue |
| Code Snippet as Input | | Convert from normal rgb to java hsb |
| Golden Reference | | Color.RGBtoHSB Color.getHSBColor |

## Appendix F. Program Synthesis

All the CodeTrans models for this task receive more than 90% accuracy, because all the samples in the tasks have very similar questions and answers in the train, validation, and test dataset. Therefore, the models could give exactly the same outputs as the reference.

| Model | Size | Model Output |
|---|---|---|
| CodeTrans Single-Task Learning | Small | [ map a [ partial1 b - ] ] |
| | Base | [ map a [ partial1 b - ] ] |
| CodeTrans Transfer Learning | Small | [ map a [ partial1 b - ] ] |
| | Base | [ map a [ partial1 b - ] ] |
| | Large | [ map a [ partial1 b - ] ] |
| CodeTrans Multi-Task Learning | Small | [ map a [ partial1 b - ] ] |
| | Base | [ map a [ partial1 b - ] ] |
| | Large | [ map a [ partial1 b - ] ] |
| CodeTrans Multi-Task Learning Fine-Tuning | Small | [ map a [ partial1 b - ] ] |
| | Base | [ map a [ partial1 b - ] ] |
| | Large | [ map a [ partial1 b - ] ] |
| Code Snippet as Input | | you are given an array of numbers a and a number b , compute the difference of elements in a and b |
| Golden Reference | | [ map a [ partial1 b - ] ] |

## References

Steven Bird. Nltk: the natural language toolkit. In *Proceedings of the COLING/ACL 2006 Interactive Presentation Sessions*, pages 69–72, 2006.

Tom B Brown, Benjamin Mann, Nick Ryder, Melanie Subbiah, Jared Kaplan, Prafulla Dhariwal, Arvind Neelakantan, Pranav Shyam, Girish Sastry, Amanda Askell, et al. Language models are few-shot learners. *arXiv preprint arXiv:2005.14165*, 2020.

Ciprian Chelba, Tomas Mikolov, Mike Schuster, Qi Ge, Thorsten Brants, Phillipp Koehn, and Tony Robinson. One billion word benchmark for measuring progress in statistical language modeling. *arXiv preprint arXiv:1312.3005*, 2013.

Kyunghyun Cho, Bart Van Merriënboer, Caglar Gulcehre, Dzmitry Bahdanau, Fethi Bougares, Holger Schwenk, and Yoshua Bengio. Learning phrase representations using rnn encoder-decoder for statistical machine translation. *arXiv preprint arXiv:1406.1078*, 2014.

Jacob Devlin, Ming-Wei Chang, Kenton Lee, and Kristina Toutanova. Bert: Pre-training of deep bidirectional transformers for language understanding. *arXiv preprint arXiv:1810.04805*, 2018.

Zhangyin Feng, Daya Guo, Duyu Tang, Nan Duan, Xiaocheng Feng, Ming Gong, Linjun Shou, Bing Qin, Ting Liu, Daxin Jiang, et al. Codebert: A pre-trained model for programming and natural languages. *arXiv preprint arXiv:2002.08155*, 2020.

Xiaodong Gu, Hongyu Zhang, Dongmei Zhang, and Sunghun Kim. Deep api learning. In *Proceedings of the 2016 24th ACM SIGSOFT International Symposium on Foundations of Software Engineering*, pages 631–642, 2016.

Sepp Hochreiter and Jürgen Schmidhuber. Long short-term memory. *Neural computation*, 9(8):1735–1780, 1997.

Xing Hu, Ge Li, Xin Xia, David Lo, and Zhi Jin. Deep code comment generation. In *2018 IEEE/ACM 26th International Conference on Program Comprehension (ICPC)*, pages 200–20010. IEEE, 2018.

Hamel Husain, Ho-Hsiang Wu, Tiferet Gazit, Miltiadis Allamanis, and Marc Brockschmidt. Codesearchnet challenge: Evaluating the state of semantic code search. *arXiv preprint arXiv:1909.09436*, 2019.

Srinivasan Iyer, Ioannis Konstas, Alvin Cheung, and Luke Zettlemoyer. Summarizing source code using a neural attention model. In *Proceedings of the 54th Annual Meeting of the Association for Computational Linguistics (Volume 1: Long Papers)*, pages 2073–2083, 2016.

Siyuan Jiang, Ameer Armaly, and Collin McMillan. Automatically generating commit messages from diffs using neural machine translation. In *2017 32nd IEEE/ACM International Conference on Automated Software Engineering (ASE)*, pages 135–146. IEEE, 2017.

Stefan Kombrink, Tomáš Mikolov, Martin Karafiát, and Lukáš Burget. Recurrent neural network based language modeling in meeting recognition. In *Twelfth annual conference of the international speech communication association*, 2011.

Taku Kudo. Subword regularization: Improving neural network translation models with multiple subword candidates. *arXiv preprint arXiv:1804.10959*, 2018.

Marie-Anne Lachaux, Baptiste Roziere, Lowik Chanussot, and Guillaume Lample. Unsupervised translation of programming languages. *arXiv preprint arXiv:2006.03511*, 2020.

Zhenzhong Lan, Mingda Chen, Sebastian Goodman, Kevin Gimpel, Piyush Sharma, and Radu Soricut. Albert: A lite bert for self-supervised learning of language representations. *arXiv preprint arXiv:1909.11942*, 2019.

Chin-Yew Lin. Rouge: A package for automatic evaluation of summaries. In *Text summarization branches out*, pages 74–81, 2004.

Chin-Yew Lin and Franz Josef Och. Orange: a method for evaluating automatic evaluation metrics for machine translation. In *COLING 2004: Proceedings of the 20th International Conference on Computational Linguistics*, pages 501–507, 2004.

Xiaodong Liu, Pengcheng He, Weizhu Chen, and Jianfeng Gao. Multi-task deep neural networks for natural language understanding. *arXiv preprint arXiv:1901.11504*, 2019a.

Yinhan Liu, Myle Ott, Naman Goyal, Jingfei Du, Mandar Joshi, Danqi Chen, Omer Levy, Mike Lewis, Luke Zettlemoyer, and Veselin Stoyanov. Roberta: A robustly optimized bert pretraining approach. *arXiv preprint arXiv:1907.11692*, 2019b.

Minh-Thang Luong, Hieu Pham, and Christopher D Manning. Effective approaches to attention-based neural machine translation. *arXiv preprint arXiv:1508.04025*, 2015.

Vadim Markovtsev and Waren Long. Public git archive: A big code dataset for all. In *Proceedings of the 15th International Conference on Mining Software Repositories*, pages 34–37, 2018.

Terence Parr. *The definitive ANTLR 4 reference*. Pragmatic Bookshelf, 2013.

Illia Polosukhin and Alexander Skidanov. Neural program search: Solving programming tasks from description and examples. *arXiv preprint arXiv:1802.04335*, 2018.

Matt Post. A call for clarity in reporting bleu scores. *arXiv preprint arXiv:1804.08771*, 2018.

Colin Raffel, Noam Shazeer, Adam Roberts, Katherine Lee, Sharan Narang, Michael Matena, Yanqi Zhou, Wei Li, and Peter J. Liu. Exploring the limits of transfer learning with a unified text-to-text transformer. *Journal of Machine Learning Research*, 21(140): 1–67, 2020. URL http://jmlr.org/papers/v21/20-074.html.

Pranav Rajpurkar, Jian Zhang, Konstantin Lopyrev, and Percy Liang. Squad: 100,000+ questions for machine comprehension of text. *arXiv preprint arXiv:1606.05250*, 2016.

Veselin Raychev, Pavol Bielik, and Martin Vechev. Probabilistic model for code with decision trees. *ACM SIGPLAN Notices*, 51(10):731–747, 2016.

Winston W Royce. Managing the development of large software systems: concepts and techniques. In *Proceedings of the 9th international conference on Software Engineering*, pages 328–338, 1987.

Sebastian Ruder. An overview of multi-task learning in deep neural networks. *arXiv preprint arXiv:1706.05098*, 2017.

Ashish Vaswani, Noam Shazeer, Niki Parmar, Jakob Uszkoreit, Llion Jones, Aidan N Gomez, Łukasz Kaiser, and Illia Polosukhin. Attention is all you need. In *Advances in neural information processing systems*, pages 5998–6008, 2017.

Cody Allen Watson. *Deep Learning in Software Engineering*. PhD thesis, The College of William and Mary, 2020.

Linting Xue, Noah Constant, Adam Roberts, Mihir Kale, Rami Al-Rfou, Aditya Siddhant, Aditya Barua, and Colin Raffel. mt5: A massively multilingual pre-trained text-to-text transformer. *arXiv preprint arXiv:2010.11934*, 2020.

Zhilin Yang, Zihang Dai, Yiming Yang, Jaime Carbonell, Russ R Salakhutdinov, and Quoc V Le. Xlnet: Generalized autoregressive pretraining for language understanding. In *Advances in neural information processing systems*, pages 5753–5763, 2019.

Ziyu Yao, Daniel S Weld, Wei-Peng Chen, and Huan Sun. Staqc: A systematically mined question-code dataset from stack overflow. In *Proceedings of the 2018 World Wide Web Conference on World Wide Web*, pages 1693–1703. International World Wide Web Conferences Steering Committee, 2018.